\documentclass[reprint,aps,prx,longbibliography,floatfix,superscriptaddress]{revtex4-2}

\usepackage{amsmath,amsfonts,amssymb,amsthm,physics,mathtools}
\usepackage{algorithm}
\usepackage[]{algorithmic}
\usepackage[caption=false]{subfig}
\usepackage{hyperref}
\usepackage{qcircuit}

\newcommand{\Cl}{\mathcal{C}\ell}
\newcommand{\Herm}{\text{Herm}}
\newcommand{\supp}{\text{supp}}

\newtheorem{Lemma}{Lemma}
\newtheorem{Theorem}{Theorem}
\newtheorem*{Theorem*}{Theorem}
\newtheorem{Corollary}{Corollary}

\begin{document}

\title{Simulating quantum computation: how many ``bits'' for ``it''?}

\author{Michael Zurel}
\affiliation{Department of Physics \& Astronomy, University of British Columbia, Vancouver, Canada}
\affiliation{Stewart Blusson Quantum Matter Institute, University of British Columbia, Vancouver, Canada}

\author{Cihan Okay}
\affiliation{Department of Mathematics, Bilkent University, Ankara, Turkey}

\author{Robert Raussendorf}
\affiliation{Stewart Blusson Quantum Matter Institute, University of British Columbia, Vancouver, Canada}
\affiliation{Institute of Theoretical Physics, Leibniz Universit{\"a}t Hannover, Hannover, Germany}

\date{\today}

\begin{abstract}
    A recently introduced classical simulation method for universal quantum computation with magic states operates by repeated sampling from probability functions [M.~Zurel~et~al. PRL~260404~(2020)]. This method is closely related to sampling algorithms based on Wigner functions, with the important distinction that Wigner functions can take negative values obstructing the sampling. Indeed, negativity in Wigner functions has been identified as a precondition for a quantum speed-up. However, in the present method of classical simulation, negativity of quasiprobability functions never arises. This model remains probabilistic for all quantum computations. In this paper, we analyze the amount of classical data that the simulation procedure must track. We find that this amount is small. Specifically, for any number $n$ of magic states, the number of bits that describe the quantum system at any given time is $2n^2+O(n)$.
\end{abstract}

\maketitle

\section{Introduction}\label{Section:Introduction}

In an article of 1989~\cite{Wheeler1989}, John Archibald Wheeler argued that quantum physics required a new perspective on reality based on information theoretic concepts. He wrote: ``No element in the description of physics shows itself as closer to primordial than the elementary quantum phenomenon, that is, the elementary device-intermediated act of posing a yes-no physical question and eliciting an answer or, in brief, the elementary act of observer-participancy. Otherwise stated, every physical quantity, every it, derives its ultimate significance from bits, binary yes-or-no indications, a conclusion we epitomize in the phrase, {\em{it from bit}}.''

A prototypical realization of this view on physics has been provided in the description of quantum computation with magic states (QCM) through the $\Lambda$ polytopes~\cite{ZurelRaussendorf2020}, in which the quantum phenomena at hand are reproduced, without any deviation or approximation, by repeated sampling processes dependent on and producing bit strings. The ``{\em{It}}'' in this case is universal quantum computation, and hence all non-relativistic quantum mechanics in finite-dimensional Hilbert spaces. The ``{\em{Bits}}'' represent the binary outcomes of Pauli measurements and the labels of the vertices of the $\Lambda$ polytopes in the repeated sampling. There are finitely many such vertices for any number $n$ of magic states.

This description of quantum computation with magic states grew out of the analysis of Wigner function negativity as a precondition for a quantum computational speedup, a research programme that started with Refs.~\cite{Galvao2005} and~\cite{VeitchEmerson2012}. Specifically, in Ref.~\cite{VeitchEmerson2012} it was established that negativity in Gross' Wigner function~\cite{Gross2006,GrossPhD} is required for a quantum speedup, under the condition that the Hilbert space dimension is odd. Analogous results were subsequently established in even dimension, specifically for rebits~\cite{DelfosseRaussendorf2015} and qubits~\cite{RaussendorfZurel2020}. However, in the end it turned out that once sufficiently general (quasi)probability functions are admitted, there is no need for any negativity at all~\cite{ZurelRaussendorf2020,ZurelHeimendahl2021}. Universal quantum computation in the magic state model can be described by repeated sampling from a generalized phase space whose points are labeled by the vertices of the $\Lambda$ polytopes. This process essentially resembles a random walk, with the complication that the transition function changes from one time step to the next and can depend on the prior sampling history.
A summary of this sampling process is given in Section~\ref{Section:Preliminaries}; see in particular Theorem~\ref{Theorem:HVM} and Algorithm~\ref{Algorithm:ClassicalSimulation}. 

The question of  interest for the present work is how much classical information the simulation of quantum computations must track, i.e., {\em{What is the length of the bit strings produced in the simulation?}} For example, if it turned out that those bit strings were very long, say exponentially long in the number $n$ of magic states, this would provide a convenient explanation for the hardness of classical simulation of universal quantum computation using $\Lambda$ polytopes. If the information storage itself is inefficient, so is the processing. However, this is not what we find. We find that the bit strings are short. Specifically, they are of length $O(n^2)$. Thus, simulation of universal quantum computation based on $\Lambda$ polytopes is a {\em{small data problem}}. The presumed hardness of this simulation must come from the computational hardness of the sampling processes involved, not from moving around large amounts of data.\footnote{Elsewhere in the quantum domain there are sampling problems which are algorithmically hard classically but for which samples can be obtained quantum mechanically, for example, boson sampling~\cite{AaronsonArkhipov2011}, random circuit sampling~\cite{BoulandFefferman2019}, and certain instantaneous quantum circuits~\cite{ShepherdBremner2009}.}

Naively, the length of the bit strings labeling the vertices of $\Lambda_n$ are upper-bounded by $\log_2(|\mathcal{V}_n|)$, with $|\mathcal{V}_n|$ the size of the generalized phase space, i.e., the number of vertices of $\Lambda_n$. To date, an upper bound and a lower bound are known for this quantity, namely
\begin{equation*}
	\frac{n^2}{2}\le\log_2(|\mathcal{V}_n|)\le 4^n n^2.
\end{equation*}
The lower bound is by Karanjai, Wallman, and Bartlett~\cite{KaranjaiBartlett2018}. The upper bound comes from the upper bound theorem of polytope theory~\cite{McMullen1970} (see Appendix~\ref{Appendix:VertexCountUpperBound} for details).

The gap between the bounds is extremely wide, consistent with both efficient and inefficient storage of the bit strings. Numerical studies suggest that the number of phase space points is growing very rapidly with $n$; for $n=1$ the number of phase space points is 8, for $n=2$ it is 22320, and for $n>2$ we don't know the precise numbers, but the estimate for $n=3$ is already huge.

The following simple insight is crucial for establishing our main result: in the QCM model, for any fixed value $n$, all quantum computations start in the {\em{same}} magic state $|M\rangle^{\otimes n}$. Therefore, the question of interest for classical simulation of QCM using $\Lambda$ polytopes is not ``What is the size of the phase space $\mathcal{V}_n$?'', but rather ``What is the size of the region of $\mathcal{V}_n$ that can be reached from the initial magic state $|M\rangle^{\otimes n}$?''. This size can be computed, and it turns out to be small. It implies a new upper bound for the length of the bit strings that is within a factor of four of the matching lower bound~\cite{KaranjaiBartlett2018}. This is the content of our main result, Theorem~\ref{Theorem:MainResult}, presented in Section~\ref{Section:MainResult} below.

\begin{figure}[b]
    \centering
    \subfloat[\label{SamplFigA}]{
        \centering
        \includegraphics[width=0.4\textwidth]{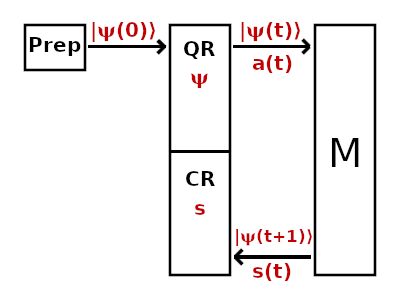}
    }
    
    \subfloat[\label{SamplFigB}]{
        \centering
        \includegraphics[width=0.4\textwidth]{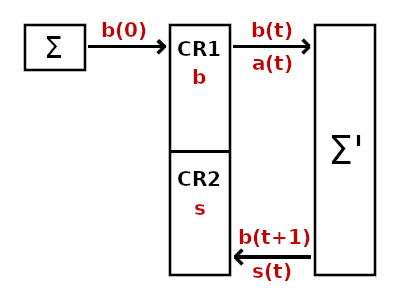}
    }
    \caption{\label{SamplFig}Quantum computation with magic states (a), and its simulation based on $\Lambda$ polytopes (b). }
\end{figure}

\section{Quantum state as a string of bits}

A crucial feature of the $\Lambda$ polytope formalism is that the quantum state $\ket{\Psi(t)}$ of the system  is replaced by a bit string $b(t)$ of bounded length. {\em{That bit string $b(t)$ is a valid and accurate representation of the quantum system.}}

The role of the bit string $b(t)$ is illustrated in Fig. \ref{SamplFig}. QCM consists of preparing a quantum register in a magic state $\ket{M}^{\otimes n}$, followed by a sequence of Pauli measurements; see Fig.~\ref{SamplFigA}. This requires a device {\sf{Prep}} to deliver the magic states to the quantum register {\sf{QR}}, and a classical register {\sf{CR}} to store the previous measurement record $\textbf{s}$, a classical side computation to identify the label $a(t)$ of the Pauli observable measured in step $t$, and a measurement device {\sf{M}} to perform the measurements and to output the corresponding results $s(t)$. The overall structure of the classical simulation is the same, but with the components modified; see Fig.~\ref{SamplFigB}. {\sf{Prep}} is replaced by a first sampler {$\mathsf{\Sigma}$} that samples from the phase space distribution of the initial state $\ket{M}^{\otimes n}$. There are two classical registers, {\sf{CR1}} and {\sf{CR2}}. The former stores the phase space samples $b(t)$, and the latter the prior measurement record, as in (a). The measurement device {\sf{M}} is replaced by a second sampler {$\mathsf{\Sigma'}$} that takes as input a phase space point $b(t)$ and a Pauli label $a(t)$, and outputs a new phase space point $b(t+1)$ as well as a measurement outcome $s(t)$. The same information that in the standard quantum mechanical description is carried by the quantum state $|\Psi(t)\rangle$ is in the $\Lambda$ polytope description carried by the bit string $b(t)$.

For universal quantum computation in the magic state model, the classical simulation needs to reproduce the quantum mechanical prediction for the joint distribution of outcomes for any sequence of Pauli measurements. The $\Lambda$-polytope method does that, without any approximation. For the predictions, the statistical distribution of the bit strings $\{b(t),\,\forall t\}$ matters, not individual values $b(t)$. This is the same for the quantum mechanical states $\ket{\Psi(t)}$. They too are conditioned on prior measurement outcomes, hence probabilistic.

\vspace{1cm}

\section{Preliminaries}\label{Section:Preliminaries}

\subsection{Quantum computation with magic states}

In Ref.~\cite{ZurelRaussendorf2020}, a hidden variable model (HVM) is defined for quantum computation with magic states (QCM)~\cite{BravyiKitaev2005}---a universal model of quantum computation in which computation proceeds through a sequence of Clifford gates and Pauli measurements on an initially prepared ``magic state''.

The measurements come from the $n$-qubit Pauli group $\mathcal{P}_n$, the group generated by the Pauli operators $X$, $Y$, $Z$ acting on $n$ qubits. Modding out overall phases we have $\mathcal{P}_n/\mathcal{Z}(\mathcal{P}_n)\cong\mathbb{Z}_2^{2n}$, where $\mathcal{Z}(\mathcal{P}_n)\cong\langle1,i,-1,-i\rangle$ is the centre of the group $\mathcal{P}_n$, and without loss of generality we can fix a phase convention for the Pauli operators to be
\begin{equation*}
	T_a=i^{-\braket{a_z}{a_x}}\bigotimes\limits_{k=1}^nZ^{a_z[k]}X^{a_x[k]},
\end{equation*}
$\forall a=(a_z,a_x)\in\mathbb{Z}_2^n\times\mathbb{Z}_2^n=:E_n$ where the inner product $\braket{a_z}{a_x}$ is computed mod $4$. The projector corresponding to a Pauli measurement $a\in E_n$ yielding outcome $s\in\mathbb{Z}_2$ is denoted $\Pi_a^s:=(1+(-1)^sT_a)/2$. The gates of the model are chosen from the Clifford group---the normalizer of the Pauli group in the unitary group up to overall phases, $\Cl_n=\mathcal{N}(\mathcal{P}_n)/U(1)$. They are defined by the property that they map Pauli operators to Pauli operators under conjugation. For more background on QCM, see Appendix~\ref{Appendix:MagicStates}.

\subsection{The \texorpdfstring{$\Lambda$}{Lambda} polytope model}
We denote by $\Herm(\mathcal{H})$ the space of Hermitian operators on Hilbert space $\mathcal{H}$, and unless otherwise specified, $\mathcal{H}$ (or $\mathcal{H}_n$) is the $n$-qubit Hilbert space $(\mathbb{C}^2)^{\otimes n}$. $\Herm_1(\mathcal{H})$ is the affine subspace of $\Herm(\mathcal{H})$ consisting of operators with unit trace and $\Herm_1^{\succeq0}(\mathcal{H})$ is the subset of $\Herm_1(\mathcal{H})$ consisting of positive semidefinite operators. $\Herm_1^{\succeq0}(\mathcal{H})$ contains the density operators representing physical quantum states.

The state space of the hidden variable model of Ref.~\cite{ZurelRaussendorf2020} is based on the $\Lambda$ polytopes. Denoting the set of pure $n$-qubit stabilizer states by $\mathcal{S}_n$, the $\Lambda$ polytope for $n$ qubits is defined as
\begin{equation}
	\Lambda_n=\left\{X\in\Herm_1(\mathcal{H}_n)\,|\,\Tr(\ket{\sigma}\bra{\sigma}X)\ge0\;\forall\ket{\sigma}\in\mathcal{S}_n\right\}.
\end{equation}
For any fixed number $n\in\mathbb{N}$ of qubits, $\Lambda_n$ is a bounded polytope with a finite number of vertices~\cite{ZurelHeimendahl2021}. We denote the vertices of $\Lambda_n$ by $\{A_\alpha\;|\;\alpha\in\mathcal{V}_n\}$ where $\mathcal{V}_n$ is an index set~\footnote{$\Lambda_n$ refers to a specific mathematical structure, a polytope, for a fixed number of qubits $n$. We refer to these polytopes collectively as the $\Lambda$ polytopes, or $\Lambda$ (no subscript). The set of vertices of $\Lambda_n$ is $\{A_\alpha\;|\;\alpha\in\mathcal{V}_n\}$. That is, $A_\alpha$ is a vertex of $\Lambda_n$ and it is labeled by $\alpha\in\mathcal{V}_n$. $\mathcal{V}_n$ is the set of all labels of vertices of $\Lambda_n$. There is no $n$ subscript on $A_\alpha$ (or $\alpha$) even though it refers to an object associated to a fixed number of qubits $n$, because this fact is implicit in the statement $\alpha\in\mathcal{V}_n$.}. The hidden variable model is defined by the following theorem.
\begin{Theorem}[Ref.~\cite{ZurelRaussendorf2020}, Theorem~1 \& Ref.~\cite{ZurelHeimendahl2021}, Theorem~1]\label{Theorem:HVM}
	For any number of qubits $n\in\mathbb{N}$,
	\begin{enumerate}
		\item Any $n$-qubit quantum state $\rho\in\Herm_1^{\succeq0}(\mathcal{H}_n)$ can be decomposed as
		\begin{equation}
			\rho=\sum\limits_{\alpha\in\mathcal{V}_n}p_\rho(\alpha)A_\alpha,
		\end{equation}
		with $p_\rho(\alpha)\ge0$ for all $\alpha\in\mathcal{V}_n$, and $\sum_\alpha p_\rho(\alpha)=1$. I.e. any $n$-qubit quantum state $\rho$ can be represented by a probability distribution $p_\rho$ over $\mathcal{V}_n$.
		\item For any $A_\alpha,\;\alpha\in\mathcal{V}_n$, and any Clifford gate $g\in\Cl_n$, $gA_\alpha g^\dagger$ is a vertex of $\Lambda_n$. This defines an action of the Clifford group on $\mathcal{V}_n$ as $gA_\alpha g^\dagger=:A_{g\cdot\alpha}$ where $g\cdot\alpha\in\mathcal{V}_n$.
		\item For any $A_\alpha,\;\alpha\in\mathcal{V}_n$, and any Pauli projector $\Pi_a^s$, we have
		\begin{equation}
			\Pi_a^sA_\alpha\Pi_a^s=\sum\limits_{\beta\in\mathcal{V}_n}q_{\alpha,a}(\beta,s)A_\beta,
		\end{equation}
		with $q_{\alpha,a}(\beta,s)\ge0$ for all $\beta\in\mathcal{V}_n$ and $s\in\mathbb{Z}_2$, and $\sum_{\beta,s}q_{\alpha,a}(\beta,s)=1$. I.e. Pauli measurements are represented by a stochastic map from (phase-space-point, measurement) pairs $(\alpha,a)\in\mathcal{V}_n\times E_n$ to (phase-space-point, measurement outcome) pairs $(\beta,s)\in\mathcal{V}_n\times\mathbb{Z}_2$.
	\end{enumerate}
\end{Theorem}

A classical simulation algorithm for QCM based on sampling from the defining probability distributions of this HVM is given in Algorithm~\ref{Algorithm:ClassicalSimulation}. The algorithm returns samples from the distribution of measurement outcomes for the quantum circuit being simulated which agree with the predictions of quantum theory~\cite{ZurelRaussendorf2020}.

\begin{algorithm}[H]
	\caption{Classical simulation of a single run of a magic state quantum circuit with input state $\rho$}
	\label{Algorithm:ClassicalSimulation}
	\begin{algorithmic}[1]
		\STATE sample $\alpha\in\mathcal{V}_n$ according to $p_\rho:\mathcal{V}_n\rightarrow\mathbb{R}_{\ge0}$
		\STATE propagate $\alpha$ through the circuit
		\WHILE{the end of the circuit has not been reached}
		\IF{a Clifford gate $g\in\Cl_n$ is encountered}
		\STATE update the phase space point according to $\alpha\leftarrowtail g\cdot\alpha$
		\ENDIF
		\IF{a Pauli measurement $a\in E_n$ is encountered}
		\STATE sample $(\beta,s)\in\mathcal{V}_n\times\mathbb{Z}_2$ according to $q_{\alpha,a}$
		\RETURN $s\in\mathbb{Z}_2$ as the outcome of the measurement
		\STATE update the phase space point according to $\alpha\leftarrowtail\beta$
		\ENDIF
		\ENDWHILE
	\end{algorithmic}
\end{algorithm}

\section{Main result}\label{Section:MainResult}

In this section we present our main result---the amount of classical data that the simulation procedure Algorithm~\ref{Algorithm:ClassicalSimulation} must track is small. First, in section~\ref{Section:MainResult1} we give the result for a simplified version of the computational model wherein we only allow Pauli measurements, no Clifford gates, and in particular, we allow only sequences of independent and commuting Pauli measurements. This simplified model is still universal for quantum computation~\cite{BravyiSmolin2016,PeresGalvao2021}. Then, in Section~\ref{Section:MainResult2} we give a more general statement of the main result where we allow computations consisting of any sequence of Clifford gates and Pauli measurements.

\subsection{Simplified case}\label{Section:MainResult1}

Although the most general quantum computation in QCM could consist of any sequence of Clifford gates and Pauli measurements performed on an arbitrary input state, we can make several assumptions simplifying the computational model while preserving the property of quantum computational universality.

First, in QCM, we usually assume that every computation starts from a fixed magic input state. For example, using the standard magic state circuit gadget~\cite[Figure~10.25]{NielsenChuang2010}, any Clifford+$T$ circuit with $n$ $T$-gates acting on $m$ qubits can be turned into a QCM circuit (Clifford gates and Pauli measurements only) on $n+m$ qubits acting on a state of the form $\ket{0}^{\otimes m}\otimes\ket{H}^{\otimes n}$ where $\ket{H}=(\ket{0}+\exp(i\pi/4)\ket{1})/\sqrt{2}$. In general, it suffices to consider input states of the form $\ket{0}^{\otimes m}\otimes\ket{M}^{\otimes n}$ where $\ket{M}$ is a fixed single-qubit magic state~\cite{Reichardt2006}.

Second, we can do away with the Clifford gates altogether~\cite{DelfosseRaussendorf2015,BravyiSmolin2016}. To see this, note that the Clifford gates can always be propagated forward in time through the circuit, conjugating the Pauli measurements into other Pauli measurements. Once they are propagated past the final measurement in the circuit they can be removed since they no longer affect the statistics of the measurement outcomes.

We can also do away with the stabilizer part of the input. The Pauli circuit on the $(m+n)$-qubit state $\ket{0}^{\otimes m}\otimes\ket{M}^{\otimes n}$ can be simulated by an adaptive sequence of Pauli measurements acting only on the $n$-qubit magic part of the input~\cite{OkayRaussendorf2021,PeresGalvao2021}. The stabilizer part is handled by extra classical side processing.

Finally, it suffices to consider only sequences of commuting Pauli measurements up to length $n$~\cite{PeresGalvao2021}. This is because when a Pauli measurement is encountered which anticommutes with a measurement that was previously performed, the outcome of the measurement will be uniformly random, and the update of the state after the measurement can be implemented by a Clifford gate. Therefore, the anticommuting measurement can always be replaced by a coin flip to determine the measurement outcome and a Clifford gate to implement the state update which can then be propagated past the future measurements. The longest sequences of independent pair-wise commuting Pauli measurements on $n$ qubits have length $n$.

To summarize, for universal quantum computation, it suffices to consider adaptive sequences of pair-wise commuting Pauli measurements of length $n$ acting on a fixed magic state of the form $\ket{M}^{\otimes n}$. A simplified version of the classical simulation algorithm above for circuits of this form is given by Algorithm~\ref{Algorithm:ReducedClassicalSimulation}. For a complete description of the above circuit simplifications, along with a method for compiling a given QCM circuit into an adaptive Pauli circuit, see Ref.~\cite{PeresGalvao2021}.

\begin{algorithm}[H]
	\caption{Classical simulation of a single run of a Pauli-based quantum circuit with input state $\rho$}
	\label{Algorithm:ReducedClassicalSimulation}
	\begin{algorithmic}[1]
		\STATE sample $\alpha_0\in\mathcal{V}_n$ according to $p_\rho:\mathcal{V}_n\rightarrow\mathbb{R}_{\ge0}$
		\FORALL{$a_t, t\in\{1,2,\dots,n\}$}
		\STATE sample $(\alpha_t,s_t)\in\mathcal{V}_n\times\mathbb{Z}_2$ according to $q_{\alpha_{t-1},a_t}$
		\RETURN $s_t\in\mathbb{Z}_2$ as the outcome of measurement $a_t$
		\ENDFOR
	\end{algorithmic}
\end{algorithm}

We can now state our main result.

\begin{Theorem}[Main result]\label{Theorem:MainResult}
	Any quantum computation consisting of a sequence of $n$ independent, pair-wise commuting Pauli measurements on a fixed magic state $\ket{M}^{\otimes n}$ can be simulated by Algorithm~\ref{Algorithm:ReducedClassicalSimulation} using a memory of $2n^2+3n$ bits to specify the phase space points reached.
\end{Theorem}

Note that the precise form of the magic input state used is not important for the proof of this theorem, so long as the state is fixed for each number of qubits. For universal quantum computation, it suffices to consider $n$ copies of a fixed single-qubit magic state such as $\ket{H}^{\otimes n}$. The proof of this theorem relies on the following result from convex geometry (see e.g. Ref.~\cite[\S1.6]{Ziegler1995}).

\begin{Theorem*}[Carath\'eodory's theorem]
	If a point $x$ of $\mathbb{R}^D$ lies in the convex hull of a set $V$, then $x$ can be written as the convex combination of at most $D+1$ points in $V$.
\end{Theorem*}

\emph{Proof of Theorem~\ref{Theorem:MainResult}.} Since the generalized phase space point operators $\{A_\alpha\,|\,\alpha\in\mathcal{V}_n\}$ are not a basis for $\Herm_1(\mathcal{H}_n)$, they are overcomplete, the distributions $p_\rho$ that represent states and the distributions $q_{\alpha,a}$ that represent Pauli measurements in the model of Theorem~\ref{Theorem:HVM} are not unique. Since $\Lambda_n$ lives in $\Herm_1(\mathcal{H}_n)$, a real affine space of dimension $4^n-1$, by Carath\'eodory's theorem there exist choices for the distribution $p_\rho$ such that $|\supp(p_\rho)|\le4^n$. Similarly, for each $s\in\mathbb{Z}_2$, there exist choices for $q_{\alpha,a}(-,s)$ such that $|\supp(q_{\alpha,a}(-,s))|\le4^n$. To start, we fix a canonical choice of the probability distributions $p_\rho$ and $q_{\alpha,a}$ satisfying these properties. With the distribution $p_\rho$ representing the input state of the circuit fixed, we can label the elements of the support by the numbers $1,2,\dots,4^n$, and so specifying a sample from this distribution (i.e., specifying an element of the support of $p_\rho$), amounts to specifying an integer between $1$ and $4^n$, which takes no more than $\log_2(4^n)=2n$ bits.

There are $4^n-1$ nontrivial $n$-qubit Pauli measurements, therefore, specifying each measurement requires no more than $2n$ bits. For the $t^{th}$ measurement $a_t$, the distribution $q_{\alpha_{t-1},a_t}$ is uniquely specified by the sampling record consisting of states $\alpha_0,\alpha_1,\dots,\alpha_{t-1}$, measurements $a_1,a_2,\dots,a_{t-1}$, and measurement outcomes $s_1,s_2,\dots,s_{t-1}$. Once this distribution is fixed, with the canonical choice above, specifying a sample from this distribution requires no more than $2n+1$ bits ($1$ bit for $s_t$, and as before $2n$ bits for $\alpha_t$).

Since the length of the measurement sequence is no more than $n$, the number of classical bits required to specify the complete sampling record is no more than 
\begin{equation*}
	\underbrace{2n}_{\alpha_0}+\sum\limits_{t=1}^n\;\bigg[\underbrace{2n}_{a_t}+\underbrace{1}_{s_t}+\underbrace{2n}_{\alpha_t}\bigg]=4n^2+3n.
\end{equation*}
See Figure~\ref{Figure:Descriptions of a computation} for a summary of the accounting used to get this bound.

This initial bound can be improved in a number of ways. First, for the purpose of simulation, we don't need to store $\alpha_n$ since there are no more measurements. This immediately removes $2n$ bits. Second, after the $t^{th}$ measurement, the value of any measurement in the span of $a_1,a_2,\dots,a_{t}$ is already determined. Therefore, if the $t+1^{th}$ measurement $a_{t+1}$ is to be independent and commute with the previous measurements, it is chosen from $\{a_1,a_2,\dots,a_t\}^\perp/\text{span}(a_1,a_2,\dots,a_{t})\cong E_{n-t}$. Specifying a measurement chosen from this set requires only $2(n-t)$ bits, not the full $2n$ bits.

Finally, we can perform another simplification which reduces the number of qubits by $1$ after each measurement. After a measurement of $a\in E_n$ giving outcome $s\in\mathbb{Z}_2$, the relevant state space is projected to $\Pi_a^s\Lambda_n\Pi_a^s$. This is contained in a $4^{n-1}-1$ dimensional subspace of $\Herm_1(\mathcal{H}_n)$. There exists a Clifford gate $g\in\Cl_n$ such that $g\Pi_{z_n}^0g^\dagger=\Pi_a^s$, and $\Pi_a^s\Lambda_n\Pi_a^s=g\Pi_{z_n}^0\Lambda_n\Pi_{z_n}^0g^\dagger=g(\Lambda_{n-1}\otimes\ket{0}\bra{0})g^\dagger$. Therefore, after the measurement the Clifford gate $g$ can be propagated out yielding a Pauli measurement circuit on an input state with a stabilizer state tensor factor. Then the above mentioned circuit simplifications~\cite{OkayRaussendorf2021,PeresGalvao2021} can be used to remove the stabilizer part of the input. This reduction can be performed after each measurement. This is similar to the idea behind the reduced classical simulation of Theorem~3 of Ref.~\cite{OkayRaussendorf2021}. With this dimension reduction after each measurement, to specify the sample $(\alpha_t,s_t)$ requires no more than $2(n-t)+1$ bits. This dimension reduction implicitly implements both of the reductions of the previous paragraph. Note that this explicit dimension reduction is not required to get the memory savings since, according to Ref.~\cite[Theorem~2]{OkayRaussendorf2021}, $\Lambda_{n-t}\otimes \ket{0}\bra{0}^{\otimes t}$ is a subpolytope of $\Lambda_n$ contained in a $4^{n-t}-1$-dimensional affine subspace of $\Herm_1(\mathcal{H}_n)$, and so by Carath\'eodory's theorem, an operator of the form $X\otimes\ket{0}\bra{0}^{\otimes t}$ can be decomposed in vertices of this subpolytope with support of size no more than $4^{n-t}$.

With these reductions, the complete measurement record can be specified with no more than
\begin{equation*}
	\underbrace{2n}_{\alpha_0}+\sum\limits_{t=1}^n\;\bigg[\underbrace{2(n-t+1)}_{a_t}+\underbrace{1}_{s_t}+\underbrace{2(n-t)}_{\alpha_t}\bigg]=2n^2+3n
\end{equation*}
classical bits, which is the claimed bound. $\Box$

\begin{figure}
	\centering
	\begin{tabular}{c c c c c c c c}
		$\rho$ & $\xlongrightarrow[]{\;\;\;a_1\;\;\;}$ & $s_1$ & $\xlongrightarrow[]{\;\;\;a_2\;\;\;}$ & $s_2$ & $\xlongrightarrow[]{\;\;\;a_3\;\;\;}$ & $s_3$ & $\cdots$\\
		& & & (a) & & & & \\&&&&&&&\\
		$p_\rho$ & $\xlongrightarrow[]{q_{\alpha_0,a_1}}$ & $(\alpha_1,s_1)$ & $\xlongrightarrow[]{a_{\alpha_1,a_2}}$ & $(\alpha_2,s_2)$ & $\xlongrightarrow[]{q_{\alpha_2,a_3}}$ & $(\alpha_3,s_3)$ & $\cdots$\\
		& & & (b) & & & & \\&&&&&&&\\
		$\alpha_0$ & $\xlongrightarrow[]{\;\;\;a_1\;\;\;}$ & $(\alpha_1,s_1)$ & $\xlongrightarrow[]{\;\;\;a_2\;\;\;}$ & $(\alpha_2,s_2)$ & $\xlongrightarrow[]{\;\;\;a_3\;\;\;}$ & $(\alpha_3,s_3)$ & $\cdots$\\
		& & & (c) & & & & \\&&&&&&&\\
		$2n$ & $+2n$ & $+2n+1$ & $+2n$ & $+2n+1$ & $+2n$ & $+2n+1$ & $\cdots$\\
		& & & (d) & & & &
	\end{tabular}
	\caption{Three descriptions of a quantum computation. (a)~A circuit level description. Pauli measurements $a_1,a_2,\dots$ are performed on the input state $\rho$ yielding measurement outcomes $s_1,s_2,\dots$. (b)~The representation of this computation in the model of Theorem~\ref{Theorem:HVM}. The input state is represented by the probability distribution $p_\rho$, each measurement $a_t$ is represented by a probability distribution $q_{\alpha_{t-1},a_t}$. (c)~A single run of the simulation algorithm based on the probabilistic representation of the computation. We start by sampling from $p_\rho$ to obtain $\alpha_0$. For each measurement $a_t$, we sample from $q_{\alpha_{t-1},a_t}$ to obtain measurement outcome $s_t$ and updated state $\alpha_t$. (d)~An upper bound on the number of classical bits required to store each piece of the sampling record of (c) according to Theorem~\ref{Theorem:MainResult}.}
	\label{Figure:Descriptions of a computation}
\end{figure}

\smallskip

A similar idea as the one used in the proof of Theorem~\ref{Theorem:MainResult} of explicitly tracking the measurement record of a quantum circuit has previously been used to define a contextual hidden variable model for the stabilizer subtheory~\cite{KaranjaiPhD}.

\subsection{General case}\label{Section:MainResult2}

Theorem~\ref{Theorem:MainResult} used a simplified version of QCM, where, without loss in computational power, all Clifford gates have been eliminated and the measurement sequence is shrunk to at most $n$ commuting Pauli measurements. In the remainder of this section we demonstrate that the bound on the size of the reachable phase space region does not increase much if we do not make these simplifications, i.e., if we admit arbitrarily long sequences of (potentially non-commuting) Pauli measurements, and Clifford gates between them. 

Specifically, we have the following additional result.

\begin{Corollary}\label{Corollary:MainResultGeneral}
	Any quantum computation consisting of an arbitrarily long sequence of Pauli measurements and Clifford unitaries, applied to a fixed magic state $\ket{M}^{\otimes n}$, can be simulated  using a memory of $4n^2+6n$ bits to specify the reachable phase space points.
\end{Corollary}
Comparing with Theorem~\ref{Theorem:MainResult} we find that the memory requirement merely doubles, in particular, the quadratic scaling with the number $n$ of magic states remains unchanged.

\smallskip

{\em{Proof of Corollary~\ref{Corollary:MainResultGeneral}.}} Every vertex of $\Lambda_n$ is in exactly one orbit of vertices with respect to the action of the Clifford group. Therefore, the number of Clifford orbits travelled by Algorithm~\ref{Algorithm:ReducedClassicalSimulation} is smaller than or at most equal to the number of vertices travelled. A consequence of Theorem~\ref{Theorem:MainResult} thus is that the number of equivalence classes travelled by Algorithm~\ref{Algorithm:ReducedClassicalSimulation} is no more than $2^{2n^2+3n}$.

At every given point in the classical simulation, the single vertex $A_\nu$ under consideration by the simulation algorithm has a product structure, $A_\nu = g (\tilde{A}_\nu \otimes \ket{S}\bra{S}) g^\dagger$, where $\ket{S}\bra{S}$ is the projector onto an $m$-qubit stabilizer state with stabilizer group $S$, $\tilde{A}_\nu$ is a vertex of $\Lambda_{n-m}$, and $g$ is a Clifford unitary~\cite{OkayRaussendorf2021}.

When switching from Algorithm~\ref{Algorithm:ReducedClassicalSimulation} specialized to the simplified computational model back to the general Algorithm~\ref{Algorithm:ClassicalSimulation}, the following additional situations may occur at any given step: (i)~the next operation is a Clifford unitary, (ii)~the next operation is the measurement of a Pauli observable $T_a$, $a\in E_n$, such that $\pm T_a \in gSg^\dagger$, and (iii)~the next operation is the measurement of a Pauli observable $T_a$ that does not commute with all elements of $gSg^\dagger$.

In case (i), the Clifford orbit $[A_\nu]$ of $A_\nu$ doesn't change. In case (ii), the measurement outcome deterministically follows from the stabilizer, and no update of $A_\nu$ occurs at all. Hence, no update of $[A_\nu]$ occurs either. In case (iii), the measurement outcome is uniformly random, and for either outcome the ensuing transformation can be replaced by a Clifford unitary. Thus we are back to case (i)---no update of $[A_\nu]$ occurs. The conclusion is that the number of Clifford orbits of vertices reachable by the more general Algorithm~\ref{Algorithm:ClassicalSimulation}
equals the number of orbits reachable by the specialized Algorithm~\ref{Algorithm:ReducedClassicalSimulation} applying to the canonical form of QCMs.

Now, the size of every Clifford orbit of vertices of $\Lambda_n$ is upper-bounded by the size $|\Cl_n|$ of the n-qubit Clifford group. Therefore, the number of vertices reachable by Algorithm~\ref{Algorithm:ClassicalSimulation} is bounded by $2^{2n^2+3n} \times |\Cl_n|$. The number of bits required to specify an $n$-qubit Clifford gate is no more than $2n^2 + 3n$. Namely, the conjugation action of a Clifford unitary on a Pauli operator is described by a binary-valued $2n \times 2n$ matrix plus a $2n$-component binary vector for the signs of Pauli operators~\cite{NielsenChuang2010}, requiring $2n\times(2n+1)$ bits. From this, the $2n^2 - n$ constraints imposed by preservation of the Pauli commutation relations need to be subtracted. Hence, to specify a vertex reachable in the simulation using Algorithm~\ref{Algorithm:ClassicalSimulation}, we specify a Clifford orbit $[A_\nu]$ and a Clifford operation $g$, which takes no more than $4n^2+6n$ bits, as claimed. $\Box$

\smallskip

We may generalize beyond the scenario of Corollary~\ref{Corollary:MainResultGeneral} by supplementing the initial magic state $\ket{H}^{\otimes n}$ with an $m$-qubit ancilla in a stabilizer state, where the stabilizer part can be kept separate and tracked independently in the simulation. As in the previous generalization, the computational power does not increase, but a broader physical setting is represented.

Then the number of bits needed depends on both the number of qubits $m$ supporting the stabilizer part of the input, and the number of magic states $n$, namely,
\begin{equation}\label{MoreGen}
    [2m^2+m] + [4n^2+6n] + [2(n+m)^2+3(n+m)].
\end{equation}
Here, the first term accounts for the stabilizer part of the state, the second term for the the magic part (this is the first version of the bound given in the proof of Theorem~\ref{Theorem:MainResult} shown in Fig.~\ref{Figure:Descriptions of a computation}), and the third term is for the $m+n$ qubit Clifford group element that needs to be tracked, as in the proof of Corollary~\ref{Corollary:MainResultGeneral}.

In deriving Eq.~(\ref{MoreGen}), we make use of the fact that a tensor product of a projector onto an $m$-qubit stabilizer state with a vertex of $\Lambda_n$ is a vertex of $\Lambda_{m+n}$~\cite[Theorem~2]{OkayRaussendorf2021}, and we use the stabilizer formalism techniques from Ref.~\cite[\S4]{OkayRaussendorf2021} to isolate update of the stabilizer part from the magic part of the state. The number of bits tracked remains quadratic in $n$ and $m$.

\section{Discussion}\label{Section:Discussion}

To summarize, in this work we have shown that the classical simulation of universal quantum computation using the $\Lambda$-polytopes~\cite{ZurelRaussendorf2020} is a small data problem. Specifically, it is shown that, with respect to the model of quantum computation with magic states, the number of bits that represent the quantum system at any stage of the simulation is quadratic in the number $n$ of the magic states. Classical simulation of quantum computation is (presumably) still hard. In the present case this hardness does not stem from shuffling around lots of data, but instead from complicated operations on little data.

For illustration, we compare the above classical simulation method to two others, one very different, one rather similar. The first method is the straightforward simulation of a quantum system obtained by choosing a specific basis of the Hilbert space at hand, mapping operators to matrices and states to vectors. The state of the system at any moment in time is now described by exponentially many complex-valued amplitudes. The bit equivalent of a complex number is a matter of numeric precision, but in any case the amount of data to be processed is large. The state update is conceptually simple, and the computational hardness derives from the size of the objects involved. This simulation method and that of Ref.~\cite{ZurelRaussendorf2020} thus represent opposite ends of the spectrum.

The second simulation method we compare to is that of sampling from Wigner functions~\cite{VeitchEmerson2012}, applicable to odd Hilbert space dimension. Here, the overall structure of the simulation is the same as in Ref.~\cite{ZurelRaussendorf2020}, i.e., repeated sampling from a phase space. There are two important differences, however. In Ref.~\cite{VeitchEmerson2012}, (i) the sampling is computationally efficient whenever it applies, but (ii) the sampling procedure does not apply to all initial magic states. Specifically, it only applies when the Wigner function~\cite{Gross2006,GrossPhD} of the initial state is positive. Indeed, this is why negativity of the Wigner function is a precondition for computational speedup. In Ref.~\cite{ZurelRaussendorf2020}, (i) the sampling is not guaranteed to be computationally efficient, but (ii) it applies to all possible initial states.

We highlight two further aspects of the simulation method~\cite{ZurelRaussendorf2020}:
\begin{itemize}
	\item In the classical simulation of quantum computation using $\Lambda$-polytopes, the description of the system's state by $2n^2+3n$ bits does not invoke any approximation. The distributions of measurement outcomes sampled from are the exact quantum-mechanical ones. Thus, the data representing the system is genuinely discrete~\footnote{Any continuous parameter dependence is relegated to the sampling probabilities in the first sampling step.}, and for this reason we regard Ref.~\cite{ZurelRaussendorf2020} as a realization of Wheeler's ``{\em{it from bit}}'' proposal.

    \item Theorem~2 can be generalized to the odd-prime-dimensional qudit case~\cite{ZurelHeimendahl2021}. In this case, the precompilation step reducing the computation to sequences of commuting Pauli measurements is also possible~\cite{Peres2023}, but again not necessary.\smallskip
\end{itemize}
The $\Lambda$-polytopes are only beginning to be explored, and they may hold many more surprises. From the perspective of quantum computation, an important question is the following. Two known efficient classical simulation methods, the stabilizer formalism \cite{Gottesman1999} and sampling from non-negative Wigner functions \cite{VeitchEmerson2012} are reproduced and extended by the $\Lambda$-polytope method, at the level of the simplest vertices of the polytopes. Where does efficient classical simulation end?

\begin{acknowledgments}
    MZ and RR thank William G. Unruh for insightful discussions. MZ and RR are funded by NSERC, in part through the Canada First Research Excellence Fund, Quantum Materials and Future Technologies Program. RR is supported by the Alexander von Humboldt Foundation. CO is supported by the US Air Force Office of Scientific Research under the award number FA9550-21-1-0002. This work was also supported by the Horizon Europe project FoQaCiA, GA no. 101070558.
\end{acknowledgments}

\bibliography{biblio}

\appendix
\onecolumngrid

\section{Upper bound on the number of vertices of \texorpdfstring{$\Lambda$}{Lambda}}\label{Appendix:VertexCountUpperBound}

\begin{Lemma}\label{Lemma:vertexCountUpperBound}
    For any number of qubits $n\in\mathbb{N}$, the number of vertices $|\mathcal{V}_n|$ of $\Lambda_n$ satisfies
    \begin{equation*}
        \log_2(|\mathcal{V}_n|)\le 1+4^{n-1}\left[n^2-n+2\log_2(10e)\right].
    \end{equation*}
\end{Lemma}

\emph{Proof of Lemma~\ref{Lemma:vertexCountUpperBound}.} According to the upper bound theorem of polytope theory~\cite{McMullen1970} (for a more detailed exposition on the upper bound theorem see Ref.~\cite[\S8.4]{Ziegler1995}), the number of facets of a $D$-dimensional polytope with $v$ vertices is bounded by the number of facets of the $D$-dimensional cyclic polytope with $v$ vertices, denoted $C(v,D)$. By duality, the number of vertices of the $\Lambda$ polytope on $n$ qubits is bounded by the number of vertices of the polar dual of the $4^n-1$-dimensional cyclic polytope with $|\mathcal{S}_n|$ vertices.

The number of vertices of the dual of $C(v, D)$ (equivalently the number of facets of $C(v,D)$) is~\cite[\S4.7]{Grunbaum2003},
\begin{equation*}
    f_{d-1}(C(v,D))=\begin{cases}
        \frac{v}{v-m}\begin{pmatrix}v-m\\m\end{pmatrix}\quad\text{for even $D=2m$},\\
        2\begin{pmatrix}v-m-1\\m\end{pmatrix}\quad\text{for odd $D=2m+1$.}
    \end{cases}
\end{equation*}
In the case of $\Lambda_n$, $D=4^n-1$, i.e. $D$ is odd and $m=2^{2n-1}-1$. The number of facets of $\Lambda_n$ is $|\mathcal{S}_n|$. Therefore,
\begin{align*}
    |\mathcal{V}_n|\le & f_{4^n-2}\left(C(|\mathcal{S}_n|,4^n-1\right)\\
    =&2\begin{pmatrix}|\mathcal{S}_n|-2^{2n-1}\\2^{2n-1}-1\end{pmatrix}\\
    \le & 2\begin{pmatrix}|\mathcal{S}_n|-2^{2n-1}\\2^{2n-1}\end{pmatrix}\\
    \le & 2\left(\frac{e(|\mathcal{S}_n|-2^{2n-1})}{2^{2n-1}}\right)^{2^{2n-1}}.
\end{align*}
In the last line we use a standard upper bound for the binomial coefficient $\begin{pmatrix}n\\k\end{pmatrix}\le\left(\frac{en}{k}\right)^k$. Then
\begin{align*}
    \log_2(|\mathcal{V}_n|)\le & 1+2^{2n-1}\left[\log_2\left(|\mathcal{S}_n|-2^{2n-1}\right)-2n+1+\log_2(e)\right]\\
    \le & 1+2^{2n-1}\left[\log_2\left(5\cdot2^{(n^2+3n)/2}\right) - 2n+1+\log_2(e)\right]\\
    \le & 1+2^{2n-1}\left[\frac{n^2+3n}{2}-2n+1+\log_2(5e)\right]\\
    = & 1+4^{n-1}\left[n^2-n+2\log_2(10e)\right].
\end{align*}
Here in the second line we use the upper bound on the number of $n$-qubits stabilizer states, $|\mathcal{S}_n|\le5\cdot2^{(n^2+3n)/2}$, from Ref.~\cite{SingalHsieh2023}. This proves the upper bound. $\Box$

\smallskip

We also have the following somewhat simpler bound.
\begin{Corollary}\label{Corollary:vertexCountUpperBound}
    For any number of qubits $n\in\mathbb{N}$, $\log_2(|\mathcal{V}_n|)\le4^nn^2$.
\end{Corollary}

\emph{Proof of Corollary~\ref{Corollary:vertexCountUpperBound}.} For $n=1$ and $n=2$, we can enumerate the vertices of $\Lambda_n$ and we find that the numbers of vertices are $8$ and $22320$ respectively. These both satisfy the bound. Further, for $n\ge2$, $1+4^{n-1}[n^2-n+2\log_2(10e)]\le4^nn^2$ and so the remaining cases follow immediately from Lemma~\ref{Lemma:vertexCountUpperBound}. $\Box$

\section{Quantum computation with magic states}\label{Appendix:MagicStates}

Quantum computation with magic states (QCM) is a universal model of quantum computation in which the stabilizer subtheory (stabilizer states, Clifford gates, and Pauli measurements) is supplemented with so-called ``magic'' (non-stabilizer) states to achieve quantum computational universality~\cite{BravyiKitaev2005}. The stabilizer subtheory is not universal for quantum computation, and circuits consisting of elements from this subtheory can be efficiently simulated by a classical computer~\cite{Gottesman1999,AaronsonGottesman2004}.

In the circuit model, universality is provided by the capacity to perform non-Clifford unitary gates. In the magic state model, on the other hand, universality is restored by the inclusion of non-stabilizer input states. Figure~\ref{Figure:MagicStateCircuit} shows a standard example of this phenomenon, a non-Clifford $T$ gate is performed, using only stabilizer operations, by injecting the magic state $\ket{H}=(\ket{0}+e^{i\pi/4}\ket{1})/\sqrt{2}$. Therefore, given access to operations in the stabilizer subtheory, as well as copies of a suitable magic state such as $\ket{H}$, any quantum computation can be performed.

For quantum computational universality, it suffices to supplement the stabilizer operations with copies of a single-qudit magic state such as $\ket{H}$. Of course, without the capacity to perform nonstabilizer operations, it is not clear that these states can be prepared. Fortunately, there exist protocols called magic state distillation protocols in which multiple noisy copies of a magic state are consumed to produce fewer copies of the magic state with higher fidelity. This approach is one of the leading candidates for scalable, fault-tolerant quantum computation~\cite{CampbellVuillot2017}.

\begin{figure}
    \centering
    \mbox{
		\Qcircuit @C=1.5em @R=1.3em {
			\lstick{\ket{H}} & \ctrl{1} & \gate{SX} \cwx[1] & \qw & \rstick{T\ket{\psi}}\\
			\lstick{\ket{\psi}} & \gate{X} & \meter
		}
	}
    \caption{Quantum circuit showing the implementation of the non-Clifford $T$ gate in the QCM framework, via the injection of the magic state $\ket{H}=(\ket{0}+e^{i\pi/4}\ket{1})/\sqrt{2}$. Since the Clifford+$T$ gate set is universal for quantum computation, this proves that, with access to suitable magic states, QCM is also universal. Circuit reproduced from \cite[Figure~10.25]{NielsenChuang2010}.}
    \label{Figure:MagicStateCircuit}
\end{figure}

For more on quantum computation with magic states, see for example~\cite[\S II]{CampbellVuillot2017} or~\cite[Ch.~10]{NielsenChuang2010}, or for a more detailed summary see~\cite[\S SM.II]{ZurelRaussendorf2020}.

\end{document}